\documentclass[twocolumn,showpacs,aps]{revtex4}
\usepackage{graphicx}
\usepackage{times}

\def\hb{\hbar}

\def\bO{{\bf \Omega}}
\def\tO{{{\tilde \Omega}}}
\def\tbO{{{\tilde {\bf \Omega}}}}
\def\ep{\varepsilon}

\def\O{\Omega}

\def\a0{\alpha_0}
\def\a{\alpha}
\def\b{\beta}
\def\g{\gamma}

\def\rtf{\rho_{\hbox{\tiny TF}}}

\def\r0{\rho_{0}}

\def\be{\begin{equation}}
\def\ee{\end{equation}}
\def\beq{\begin{equation}}
\def\eeq{\end{equation}}

\def\cd{{\cal D}}

\begin{document}

\title{Three-dimensional vortex configurations in a rotating Bose Einstein condensate}

\author{Amandine Aftalion}
\email{aftalion@ann.jussieu.fr} \affiliation{Laboratoire
Jacques-Louis Lions,  Universit{\'e} Paris 6, 175 rue du Chevaleret,
75013 Paris, France.}
\author{Ionut Danaila}
\email{danaila@ann.jussieu.fr} \affiliation{Laboratoire
Jacques-Louis Lions,  Universit{\'e} Paris 6, 175 rue du Chevaleret,
75013 Paris, France.}
\date{\today}

\pacs{03.75.Fi,02.70.-c}

\begin{abstract}
We consider a rotating Bose-Einstein condensate in a harmonic trap
and investigate numerically the behavior of the wave function
which solves the Gross Pitaevskii equation.
 Following recent experiments \cite{RBD}, we study in detail the line of a single
quantized vortex, which has a $U$ or $S$ shape. We find that a
single vortex can lie only in the $x-z$ or $y-z$ plane. $S$ type
vortices exist for all values of the angular velocity ${\O}$ while
$U$ vortices exist for ${\O}$ sufficiently large.  We compute the
energy of the various configurations with several vortices and
study the three-dimensional structure of vortices.
\end{abstract}

\maketitle

\section{Introduction}

Several experimental groups have produced vortices in Bose
Einstein condensates (BEC) \cite{mal,AK,MCWD,MCWD2,RK,RBD}. One
type of experiments  consists in imposing a laser beam on the
magnetic trap holding the atoms to create a harmonic anisotropic
rotating potential. For a
prolate trap,  it has been observed \cite{MCWD2,MCWD,RBD} that
when a single vortex exists,  the vortex line is not straight
along the axis of rotation, but bending. Theoretical works
\cite{AR,AJ} establish a simpler expression of the
Gross Pitaevskii energy that only depends on the vortex
lines. In \cite{AJ},  it is proved that bending occurs for
prolate condensates, but not for oblate ones.

Minimization algorithms \cite{GP1,CM} have been used to compute local
minima of the Gross Pitaevskii  energy and provide an evidence of
the bending in the same setting as the experiment. Bending (or
$U$) vortices are described in detail, and multiple vortex
configurations are addressed in these studies.

Recently, the ENS group \cite{RBD} has further studied
configurations with a single vortex line. They have observed
planar bent vortices ($U$) but also different
configurations ($S$). They study the length of the
line, its deviation from the center and its angular momentum.

In this paper, motivated by the recent experiments at the ENS
\cite{RBD}, we numerically look for local minimizers of the Gross
Pitaevskii energy and we want to understand the various vortex
configurations observed in the experimental setting: $U$ vortices
but also $S$ vortices. We compute solutions with up to 4 vortices
and describe their three-dimensional structure. Different solution
branches are followed and the evolution of the
corresponding energy and angular momentum are shown. The framework
of this study is the case of a prolate condensate where bending is
an important phenomenon.

 We consider a pure BEC of $N$ atoms confined in a harmonic
 trapping potential rotating along the $z$ axis at angular
 velocity ${\O}$. The equilibrium of the system corresponds to local
 minima of the Gross-Pitaevskii energy in the rotating frame

\begin{eqnarray}
\label{BE} {\cal E}(\phi) &=& \int_{\cal D} {\hb^2 \over{2m}}
|\nabla \phi|^2+{{\hb \bO}}\cdot (i\phi, \nabla \phi\times {\bf
x}) \nonumber\\ \nonumber &&+{m\over 2}
  \omega^2_x  (x^2+\a^2 y^2+\b^2z^2) |\phi|^2 +{N}
  g_{3D}|\phi|^4,
\end{eqnarray}
where $g_{3D}=4\pi\hb ^2a/m$ and the wave function $\phi$ is
normalized to unity $\int_{\cal D} |\phi|^2 =1$. Here, for any
complex quantities $u$ and $v$ and their complex conjugates
$\bar{u}$ and $\bar{v}$, $(u,v)=(u\bar{v}+\bar{u}v)/2$.

For numerical applications, it is more convenient to rescale the
variables  as follows: ${\bf r}={\bf x}/R$, $ u({\bf r})=R^{3/2}
\phi({\bf x})$, where $R=d/\sqrt{\ep}$ and
$$d=\left(\frac{\hb}{m\omega_x}\right)^{1/2}, \,\, \ep=\left({{d}\over {8\pi
Na}}\right)^{2/5}, \,\, \tO ={\O}/{(\ep\omega_x)}.$$ In this
scaling the Thomas-Fermi limit of $u$ is \beq\label{rtf} \rtf({\bf
r})=\r0 -(x^2+\a^2y^2+\b^2z^2).\eeq Then, we use the
dimensionless energy introduced in \cite{AR} \beq \label{be3d}
{E}(u) = H(u) - \tO L_z(u),\eeq with \beq\label{H}
H(u)=\int_{\cd}{1\over 2} |\nabla u|^2 -{1\over
2\ep^2}\rtf|u|^2+{1\over 4\ep^2}|u|^4\;,\eeq \beq\label{Lz}
L_z(u)=i \int_{\cd} \bar{u} \left( y\frac{\partial u}{\partial x}
- x \frac{\partial u}{\partial y}\right), \eeq defined in  the
domain ${\cal D}=\left\{\rtf({\bf r}) \geq 0\right\}.$

\subsection{Numerical method}

In the present study we compute critical points of ${E}(u)$
by solving the norm-preserving imaginary time propagation of the
corresponding equation: \beq\label{eqstrong} \frac{\partial
u}{\partial t}-{1\over 2}\Delta u + i(\tbO\times {\bf r}).\nabla
u={1\over {2\ep^2}}u(\rtf-|u|^2)+\mu_\ep u, \eeq with $u=0$ on
$\partial \cd$ and $\mu_\ep$ the Lagrange multiplier for the norm
constraint $\int_{\cal D} |u|^2=1$. A hybrid 3 steps
Runge-Kutta-Crank-Nicolson  scheme \cite{Orlandi} is used to
advance the equation in time:
\begin{equation}
  \frac{u_{l+1}-u_l}{\Delta t}=a_l {\cal H}_l + b_l {\cal H}_{l-1}
  +c_l \Delta \left( \frac{u_{l+1}+u_l}{2}\right),
\end{equation}
where ${\cal H}$ contains the remaining non-linear terms. The
corresponding constants for every step ($l=1,2,3$) are :
\[
\begin{array}{lll}
a_1=8/15,& a_2=5/12,& a_3=3/4,\\
   b_1=0,& b_2=-17/60,& b_3=-5/12,\\
      c_1=8/15,& c_2=2/15,& c_3=1/3.
\end{array}
\]
The resulting semi-implicit scheme is second order time accurate
and allows reasonably large time steps, making it appropriate for
long time integration. The large sparse matrix linear systems
resulting from the implicit terms are solved by an alternating
direction implicit (ADI) factorization technique.

For the spatial discretization we use finite differences on a
Cartesian uniform mesh with periodic boundary conditions in all
directions. To accurately resolve sharp gradients of the variable
in presence of vortices, low numerical dissipation and very
accurate schemes are required for the spatial derivatives. A
sixth-order compact finite difference scheme \cite{Lele} with
spectral-like resolution was chosen to this end.

\subsection{Physical and numerical parameters}\label{sec-param}

The values of constants in (\ref{eqstrong}) are set to $\ep=0.02,
\quad \a=1.06,\quad \beta=0.067,$ corresponding to the experiments
of the ENS group \cite{CM,MCWD} ($m=1.445\cdot10^{-26}kg$,
$a=5.8\cdot10^{-11}m$, $N=1.4\cdot10^{5}$ and
$\omega_x=1094s^{-1}$). The angular frequency ${\O}$ will be varied
from 0 to the maximum value of $0.9 \omega_x$, for which no
deformation of the condensate has to be taken into account.

Equation (\ref{eqstrong}) is propagated in imaginary time until
the  evolution of the energy (\ref{be3d}) has a gradient in time smaller than
$10^{-6}$. For the considered range of $\Omega$, the numerical
domain is fixed to an elongated box $(x,y,z)\in
[-0.6,0.6]\times[-0.6,0.6]\times[-8.5, 8.5]$. A refined grid using
$72\times 72 \times 510$ nodes is employed, which  is sufficient
to achieve grid-independence for all considered numerical
experiments.

Different initial conditions are used in to trigger single or
multiple vortex configurations and follow the corresponding
branches as ${\O}$ is varied. The simplest initial condition assumes
a steady-state solution  $u(x,y,z)=\sqrt{\rtf(x,y,z)}$ and is
useful to study vortex-free configurations and their degeneracy
into multiple vortex configurations when increasing the value of
$\Omega$. Initial conditions with vortices are obtained by
superimposing to the steady-state a simple ansatz for the vortex.
For example, an initial condition with a centered straight vortex
of radius $\ep$ is obtained by imposing
\begin{eqnarray}
 u(x,y,z)=\sqrt{\rtf}\cdot u_{\ep}, \\
\nonumber u_{\ep}=\sqrt{0.5
\left\{1+tanh\left[\frac{4}{\ep}\left(r-\ep\right)\right]\right\}}
\cdot exp(i \varphi),
\end{eqnarray}
where $(r,\varphi)$ are the polar coordinates in the $(x,y)$
plane. The 3D shape of the vortex can be easily modified by
shifting the center $r_0$ of the vortex in successive $(x,y)$
planes; for instance, to obtain a planar S shape vortex,  the
following function can be used:
\[
  r_0(z)=\left\{
  \begin{array}{ll}\vspace{0.2cm}
-1 + tanh\left[\alpha_v
\left(1+{\displaystyle z\over \beta_v}\right)\right]\Big/tanh(\alpha_v), & z < 0 \\
1 + tanh\left[\alpha_v \left(-1+{\displaystyle z\over
\beta_v}\right)\right]\Big/tanh(\alpha_v), & z \geq 0
  \end{array}
\right.
\]
The constants $\alpha_v, \beta_v$ control, respectively, the
curvature and the height of the vortex.

We first focus on single vortex configurations and describe later
multi vortex configurations.

\section{Single vortex lines}

\begin{figure}[!h]
\centerline{\includegraphics[width=0.90\columnwidth]{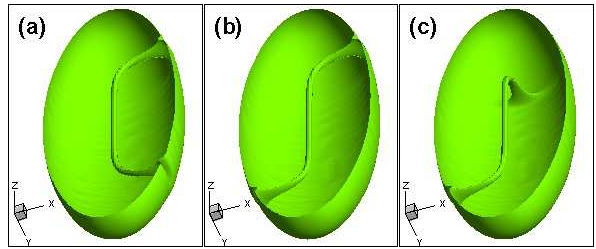}}
 \caption{Single vortex configurations in BEC:
 (a) {\em U} vortex, (b) planar {\em S} vortex, (c) non-planar  {\em S} vortex.
 Iso-surfaces of lowest density within the condensate.} \label{fig-1v-all}
\end{figure}

 We have observed three different
types of single vortex configurations as shown in figure
\ref{fig-1v-all}: planar $U$ vortices, planar $S$ vortices and
non-planar $S$ vortices. The $U$ vortices are the bent vortices
computed in \cite{GP1,CM} and theoretically studied in
\cite{AR,AJ}. They are global  minimizers of the energy.
 The $S$ configurations were observed experimentally very
recently \cite{RBD}  and are only local minimizers of the energy.

\subsection{$U$ vortex }

\begin{figure}[!h]
%\centerline{\includegraphics[width=0.90\columnwidth]{U1new_om-21-29-39.ps}}
\centerline{\includegraphics[width=0.90\columnwidth]{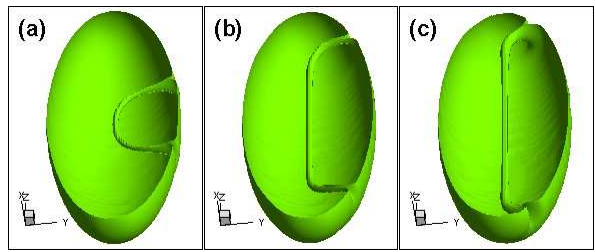}}
 \caption{Single $U$ vortex configurations for ${\O}/\omega_x=0.42$ (a), $0.58$ (b), $0.78$ (c).}
 \label{fig-U1-omega}
\end{figure}

\begin{figure}[!h]
%\centerline{\includegraphics[width=0.90\columnwidth]{U1S1-Energy.ps}}
\centerline{\includegraphics[width=0.90\columnwidth]{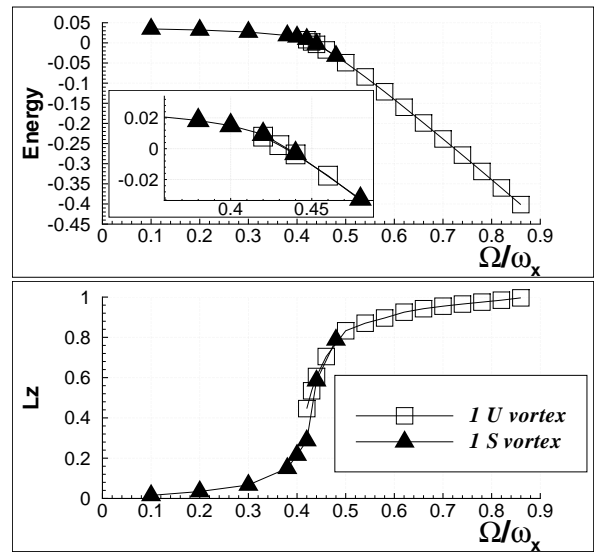}}
 \caption{Energy (in units of $\hbar \omega_x$) and angular momentum
 per particle (in units of $\hbar$) for the single vortex configurations.} \label{fig-E1v}
\end{figure}

The $U$ vortex is a planar vortex formed of 2
parts: the central part is a line which stays
on the $z$ axis and the outer part  reaches the
boundary  of the condensate perpendicularly.
When ${\O}$ increases, the central straight part
gets longer (figure \ref{fig-U1-omega}) and the
angular momentum ($L_z$) increases to 1 (figure
\ref {fig-E1v}).

The $U$ vortex is  obtained by starting the simulation with an
initial condition containing a straight vortex away from the $z$
axis. In fact, the $U$ vortex lies either in the $x-z$  or  $y-z$
plane. Starting with an initial condition which is not in one of
these plane yields a final state in  the $y-z$ plane, which is the
plane closest to the $z$ axis.

The shape of the the $U$ vortex and its preferred location in the
$y-z$ plane can be analyzed using the approximate energy derived
in \cite{AR,AJ}: setting the vortex free solution to 0 energy,
then the energy of a vortex line $\g$ can be approximated by
\beq\label{env} {\cal E}_\gamma=\int_\g \rtf\ dl-C{\O} \int_\g
\rtf^2 \ dz, \eeq where $C$ is a constant which depends on the
experimental parameters and $\rtf$ is given by (\ref{rtf}). If
$\g$ is not in the $x-z$ or $y-z$ plane, then one can construct
small perturbations of $\g$ that preserve $\rtf$ and lower the
energy. This implies that $\g$ cannot be a critical point of the
energy because the gradient is not zero. Of
course, if the ellipticity of the cross section is small, the
gradient is small, which may allow to observe these
configurations.

 In order to understand the existence of the straight central
part of the $U$ vortex, one can also refer to the analysis of
\cite{AJ}: from equation
 (\ref{env}) we can infer that  a vortex line with a lower energy
 than the vortex free solution is obtained when the quantity
 $\rtf-C{\O} \rtf^2$ is negative, {\em i.e.} $C{\O} \rtf>1$.
Let $\bar{\O}$ be such that $C\bar{\O} \rho_0=1$; it corresponds to the
2d critical velocity for the existence of a vortex in the
 plane $z=0$.  For ${\O}$ close to  $\bar
 {\O}$, the inner region where $C {\O}\rtf >1$ is concentrated near
 the center of the condensate. In this region, the vortex  line
 has to be straight (see \cite{AJ}).  This straight
 part is getting longer as ${\O}$ increases since the region where $C{\O}
 \rtf > 1$ is getting bigger.
  This region corresponds to ${\O} >
 {\O}_{2d}(z)$, where ${\O}_{2d}(z)$ is the critical velocity for the
 existence of a vortex in the 2 dimensional section where $z$ is
 constant. In the outer region,  the
 vortex reaches the boundary using the shortest path.

Figure \ref{fig-E1v} shows the energy and angular momentum
variation with $\Omega$ for the single vortex configurations. The
$U$ vortices exist only for ${\O}$ bigger than a critical value
${\O}_c=0.42 \omega_x$. It is interesting to note that at ${\O}_c$,
the energy of the $U$ vortex is bigger than the energy of the
vortex free solution (we have set to zero the energy of the vortex
free solution). A zoom in this region shows that ${\O}_c$ is
very close to
 the angular velocity ${\O}_1$ for which the energy of
the vortex free solution is equal to the energy of the $U$ vortex.
 Figure \ref{fig-E1v} also shows that the angular momentum $L_z$ of the
$U$ vortex for ${\O}={\O}_c$ does not go to 0. This suggests that in
fact there could be another $U$ solution
 for ${\O}>{\O}_c$.  Using an ansatz, another type of $U$ solution is obtained in
 \cite{CM} which is a saddle point of the energy: it is away from
 the axis and has lower angular momentum. In \cite{AJ}, it is
 proved rigorously that for ${\O}$ small, there is no $U$ as a
 critical point of the energy.

For an initial
condition with a straight vortex centered on the $z$ axis,
 if ${\O} <0.8 \omega_x$, the straight vortex is unstable and the
final configuration is a $U$, but if  ${\O}
>0.8 \omega_x$, the straight vortex is stable. This is in agreement with the result of \cite{AJ}
where the local stability of the straight vortex for   ${\O}$ larger
is proved.

For small ${\O}$, the $U$ vortex disappears and a
vortex-free configuration is obtained, while for ${\O}$ larger
the $U$ vortex degenerates into a three-vortex configuration
(described later).

\subsection{$S$ vortex}

 Motivated by the experiments of \cite{RBD}, we compute new
critical points of the energy, which are $S$ configurations (see
figure \ref{fig-1v-all}).  Several numerical experiments were
performed, starting from different initial conditions containing
an ansatz for the $S$ vortex (see section \ref{sec-param}).

\begin{figure}[!h]
\centerline{\includegraphics[width=0.75\columnwidth]{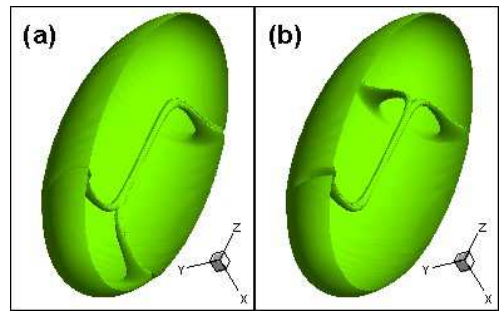}}
 \caption{Comparison between the single vortex configurations obtained for
 the same angular velocity ${\O}/\omega_x=0.44$. Superposition of
 the $U$ and $S$ vortex (a) and the planar and non-planar $S$ vortex (b).} \label{fig-U1S1}
\end{figure}

The planar $S$ can be regarded as a $U$, with the half-part in the
plane $z<0$ rotated with respect to the $z$ axis by 180 degrees
(see figure \ref{fig-U1S1}). The non planar $S$ are such  that the
projections of the branches on the $x-y$ plane are orthogonal,
{\em i.e.} the rotation of the branches is of 90 degrees.  We
could check that non planar $S$ configurations with an angle
between the branches different from 90 degrees do not exist.

As already mentioned for the $U$ vortex, stable planar $S$
configurations lie either in the $x-z$ or $y-z$ plane. As for the
$U$, this can be explained using the limiting energy obtained in
\cite{AJ} and considering separately the upper or lower part of
the $S$. As soon as the cross section is not a disc, if the upper
or lower branch of the $S$ configuration does not lie in the $x-z$
or $y-z$ plane, then the gradient of vortex line energy
(\ref{env})  can never be zero when $\g$ is varied.

\begin{figure}[!h]
%\centerline{\includegraphics[width=0.90\columnwidth]{S1new_om-19-22-24.ps}}
\centerline{\includegraphics[width=0.90\columnwidth]{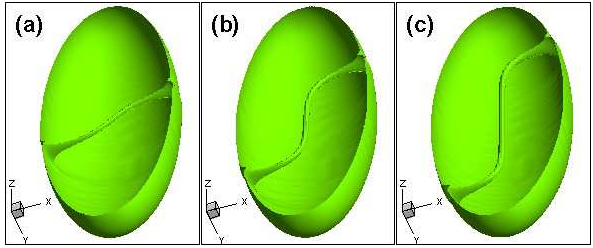}}
 \caption{Single $S$ vortex configuration for ${\O}/\omega_x=0.38$ (a), $0.44$ (b), $0.48$ (c).} \label{fig-S1-omega}
\end{figure}

The $S$ vortices exist for all values of ${\O}$ while the $U$ only
exist for ${\O}>{\O}_c$. When ${\O}$ decreases, the extension  of the
$S$ along the $z$ axis goes downwards as shown in figure
\ref{fig-S1-omega}, the angular momentum decreases to 0 (figure
\ref{fig-E1v}) and the vortex tends to the horizontal axis. Note
that
 a vortex along the horizontal axis has $L_z=0$, but a  positive energy. On the
other side, when ${\O}$ increases, the $S$ gets straighter and it
tends to the vertical axis.

The global minimum of the energy is  never an $S$. But the
difference in energy (and angular momentum) between $U$ and $S$ vortices is
very small,  as illustrated in figure \ref{fig-E1v} because an
$S$ vortex is almost like a $U$ with a half-part rotated by 180 degrees.

\subsection{Minimizer with fixed $L$}

As pointed out in \cite{RBD}, the minimization problem which is
related to the experiments, is rather to minimize $H$ (see
(\ref{H}) while fixing $L_z$, rather than minimizing $E=H-{\O} L_z$.
 This has been studied in the 2 dimensionnal setting in\cite{BS}.
 One can notice that if a given configuration with $H=h$ and
$L_z=l$ minimizes $E=H-{\O} L_z$ for some ${\O}$, then $h$ minimizes
$H$ under the constraint that $L_z=l$: indeed if $H'=H(u)$ with
$L_z(u)=l$, then $H'-{\O} l\geq h-{\O} l$, since $(h,l)$ minimizes
$E$, and this implies that $H'\geq h$. Moreover ${\O}$ is the slope
to the curve $H(L_z)$ at the point $(h,l)$ and the property of
minimizing $E$ that is for all $h'$, $l'$,
$$h'-{\O} l' \geq h-{\O} l$$
implies that the curve $H(L_z)$ lies above its tangent at this point.

We have plotted $H$ as a function of $L_z$. We can check that the
curve is convex, and above its tangent, which is consistent with
the fact that we have computed minimizers of the energy.

\begin{figure}[!h]
\centerline{\includegraphics[width=0.75\columnwidth]{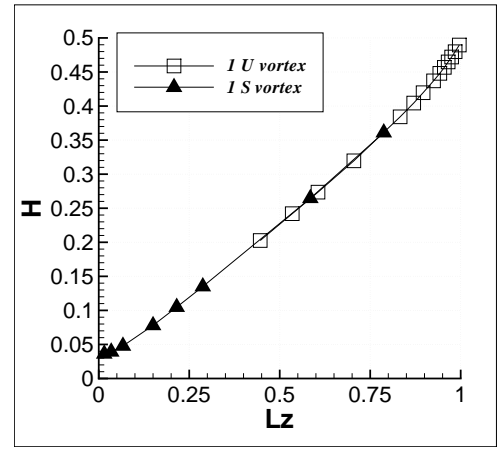}}
 \caption{Single vortex configuration.} \label{fig-LZHU}
\end{figure}

We know that the $U$ solution exists for ${\O}\geq {\O}_c$ and has
$L_z>0.4$. For $L_z<0.4$, we expect that the process of minimizing
$H$ with fixed $L_z$ would produce $U$ vortices and the curve
$H(L_z)$ should be concave in this region. In \cite{AJ}, we have
proved that for $L_z$ close to 0, $H\geq C L_z^{2/3}$, which is a
first indication to the concavity of the curve.

\section{Multiple vortices}

 Multiple
vortex configurations are obtained based upon different numerical
strategies. The first one is to start the computation from a
vortex-free steady state  and to abruptly increase $\Omega$ to a
very hight value; multiple vortices are thus obtained. The second
strategy is to generate an initial condition with vortices as
described in section \ref{sec-param} (the advantage being the
control of the shape and initial arrangement of the vortices).

\begin{figure}[!h]
\centerline{\includegraphics[width=0.75\columnwidth]{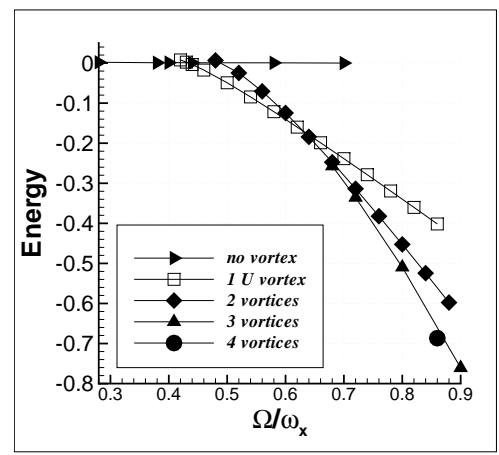}}
 \caption{Energy (in units of $\hbar \omega_x$) for all  studied configurations.} \label{fig-Eall}
\end{figure}

\begin{figure}[!h]
\centerline{\includegraphics[width=0.75\columnwidth]{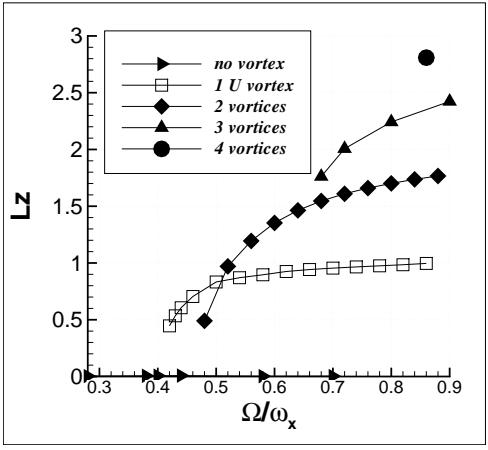}}
 \caption{Angular momentum $L_z$ (in units of $\hbar$) for all  studied configurations.} \label{fig-LZall}
\end{figure}

Both techniques are used to follow solution branches with two,
three or four vortices in the condensate. Figures \ref{fig-Eall}
and \ref{fig-LZall} display energy and angular momentum vs
$\Omega$ for all studied configurations.

\subsection{3 vortices}

\begin{figure}[!h]
\centerline{\includegraphics[width=0.90\columnwidth]{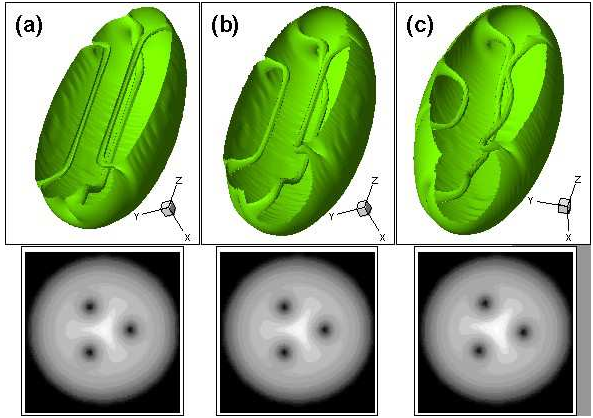}}
 \caption{Three-vortex configuration for ${\O}/\omega_x=0.9$ (a), $0.72$ (b), $0.68$ (c).
 Lower pictures show iso-contours of $|u|$ in the central $z=0$ cut plane. } \label{fig-U3-omega}
\end{figure}

When ${\O}$ is increased, the single $U$ vortex solution switches to
a 3 vortex configuration ($\Omega=0.9 \omega_x$). As shown in
figure \ref{fig-U3-omega}a, the configuration is invariant under
rotation in a central plane near $z=0$ but not near the edges. For
large $\Omega$, three-dimensional views show (figure
\ref{fig-U3-omega} a,b) that there are 2 vortices of similar size
and a longer  one which is bending near the boundary. For
$\Omega=0.8 \omega_x$, all vortices display contorted shapes
(figure \ref{fig-U3-omega}c), very similar to those reported in
\cite{GP1}. Let us point out that the angular momentum of all
these 3 vortex configurations is lower than 3 (see figure
\ref{fig-LZall}).

\begin{figure}[!h]
\centerline{\includegraphics[width=0.75\columnwidth]{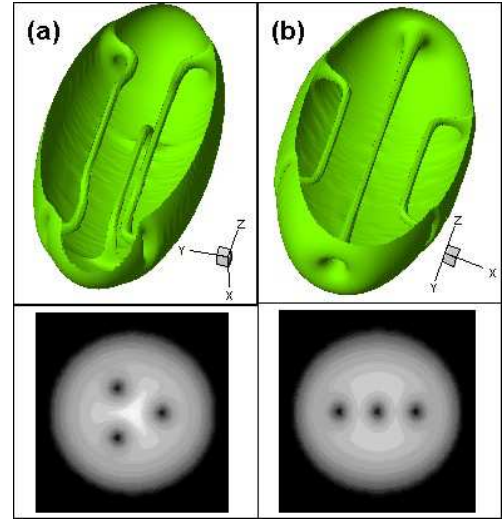}}
 \caption{Three-vortex configuration obtained for the same  ${\O}/\omega_x=0.72$,
 from different initial conditions: 3 identical $U$ vortices at 120$^o$ (a) and
 3 straight vortices in a row on the $x$-axis (b).
 Lower pictures show iso-contours of $|u|$ in the central $z=0$ cut plane. } \label{fig-U3-ligne}
\end{figure}

 When we put as initial condition a configuration with
3 identical $U$ vortices at 120$^o$, in the final state, one of
them gets a little longer (figure \ref{fig-U3-ligne}a) and  the
symmetry is lost. This configuration has almost the same energy
and angular momentum as the configuration displayed in figure
\ref{fig-U3-omega}b. In exchange, the initial condition with three
straight vortices on the $x$-axis has its symmetry preserved
(figure \ref{fig-U3-ligne} b), but with a higher energy that the
previous one.

\begin{figure}[!h]
\centerline{\includegraphics[width=0.75\columnwidth]{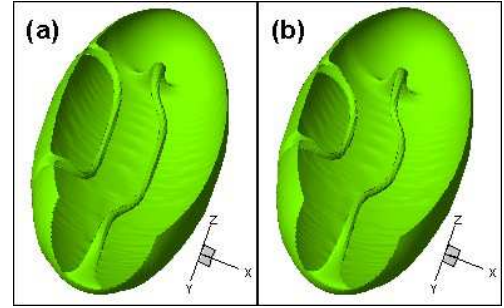}}
 \caption{Two vortices obtained from the 3-vortex configuration when the value of ${\O}/\omega_x$
 is decreased to $0.64$ (a) and $0.6$ (b).} \label{fig-U2-fromU3}
\end{figure}

When further decreasing ${\O}$, the 3 vortex branch switches to a
2-vortex displaying irregular shapes (figure \ref{fig-U2-fromU3}).

%These observations tend to confirm the results in \cite{GP1} conjecturing that only multiple vortices lacking rotational symmetry  are stable configurations in the prolate condensate. We shall see in the next section that this is not always the case for other configurations.

\subsection{2 vortices}

The two-vortex branch presented in this section was obtained by
starting from a vortex-free solution and suddenly increasing
$\Omega$ to a value of $0.8 \omega_x$. The configuration is planar
and symmetric, like twice a single $U$ vortex, but away from the
axis (there is a repulsion between the lines).

When ${\O}$ increases, the lines are almost straight and get closer to each other. This is in
agreement with the fact that when ${\O}$ gets large, the straight
vortex is a local minimizer of the energy. Hence the bending is no
longer the important phenomenon.

\begin{figure}[!h]
%\centerline{\includegraphics[width=0.90\columnwidth]{U2new_om-20-30-40.ps}}
\centerline{\includegraphics[width=0.90\columnwidth]{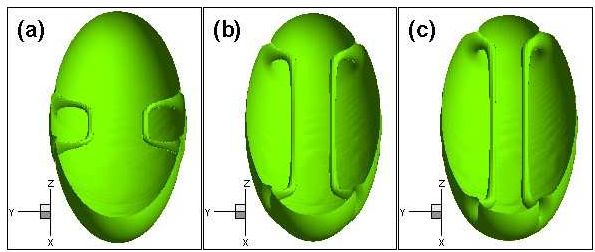}}
 \caption{Configuration with two symmetric vortices
 for ${\O}/\omega_x=0.48$ (a), $0.6$, (b) $0.8$ (c).} \label{fig-U2-omega}
\end{figure}

We recall that decreasing ${\O}$ from a configuration with 3
vortices, we obtained 2 vortices which are not symmetric, one
being longer than the other (figure \ref{fig-U2-fromU3}). This
configuration has slightly bigger energy than the 2 symmetric
vortices.

\subsection{4 vortices}

\begin{figure}[ht]
\centerline{\includegraphics[width=0.75\columnwidth]{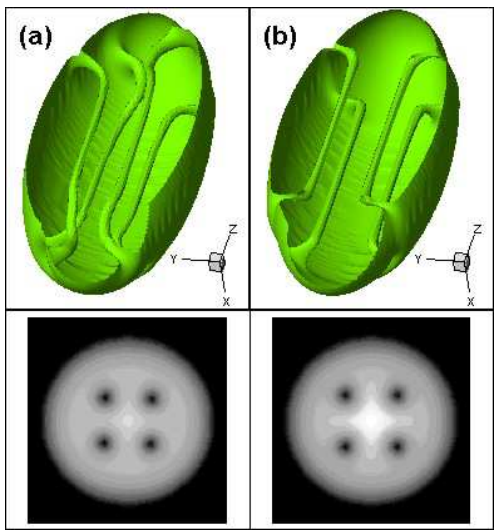}}
 \caption{Four-vortex configurations for (a) ${\O}/\omega_x=0.86$ - obtained from an
 initial condition without vortices and (b) ${\O}/\omega_x=0.72$ -
 obtained from an initial condition with four symmetrical
 vortices.} \label{fig-U4}
\end{figure}

Starting from an initial condition without vortices and increasing
$\Omega$ to $0.86\omega_x$, we have obtained stable configurations
with 4 curved vortices (figure \ref{fig-U4} a). When decreasing
$\Omega$, this configuration rapidly degenerates into a
three-vortex state. For lower $\Omega$ we could obtain stable
configurations with four symmetric vortices (figure \ref{fig-U4}
b), but with bigger energy. The location of the vortices in the
plane $z=0$ is the same.

We have to point out that with the initial condition of 4
identical vortices, the symmetry is preserved as displayed in
figure \ref{fig-U4}b, which was not the case for 3 vortices.

\section{Concluding remarks}

 We have studied different vortex configurations in a prolate Bose
Enstein condensate by solving the Gross Pitaevskii equation. We
have computed $U$ and $S$ vortices, motivated by the recent
experiments of \cite{RBD}. Our computations involve a
  parameter  $\ep$, which is small when the number of atoms $N$ is
large. Decreasing $\ep$, that is increasing the number of atoms
 forces the vortex lines to be almost straight in their central
part, while for $\ep$ larger, the central straight part is not so
obvious as in some figures of \cite{GP1}.

 We have found that the $S$ vortices are only local minimizers of the energy and exist
for all values of the angular velocity ${\O}$, while $U$ vortices
are global minimizers existing for ${\O} \geq {\O}_c$.
%(??The energy of the $U$ vortex at ${\O}_c$ is higher than the
%energy of the free vortex configuration, which confirms
%theoretical predictions of ... The angular momentum of the $S$
%vortex goes to 0 with ${\O}$, which is not the case for the $U$
%vortex at ${\O}_c$. - {\`a} mettre ou pas ?)
 A planar $S$ vortex can be regarded as a $U$ vortex with a
half-part rotated by 180$^o$. Moreover,  $U$ or planar $S$ vortices lie
only in the $x-z$ or $y-z$ plane while non planar $S$
vortices exists only for an angle of 90$^o$ between the two
branches.

We have followed the branches of solutions when varying ${\O}$ and found configurations with  two, three and four
vortices.
%When solution branches are followed by continuity, stable configurations with contorted shapes are obtained as in \cite{GP1}. Starting from artificial initial conditions with symmetrical vortices in the condensate we could obtain stable configurations conserving the symmetry for two and four vortices, but not for three vortices.

\end{document}